# Third Order Perturbed Energy of Cobalt Ferrite Thick Films

P. Samarasekara

Department of Physics, University of Peradeniya, Peradeniya, Sri Lanka.

**Abstract**

Magnetic properties and easy axis orientation of cobalt ferrite films with applied magnetic field and number of layers were studied. According to our theoretical studies explained in this manuscript, the magnetically easy and hard directions of cobalt ferrite films solely depend on in plane and out of plane magnetic fields. According to 3-D and 2-D plots, there are many easy and hard directions at one particular value of in plane or out of plane magnetic field. The magnetic properties were investigated for cobalt ferrite films with thickness up to 10,000 unit cells. The total magnetic energy was calculated for a unit spin of cobalt.

## 1. Introduction

Cobalt ferrite with inverse spinel cubic structure is a soft non-uniaxial ferrimagnetic material. In the applications of magnetic memory devices and microwave applications, where a small magnetic anisotropy is required, cobalt ferrite is used. Easy axis of cobalt ferrite is along the one of the edge of the cubic cell. Because almost all of the ferrites are oxides, they are corrosion resistive and mechanically hard. Films of cobalt ferrite have been experimentally synthesized by rf sputtering [1, 2], evaporation method [3] and pulsed laser deposition [4, 5]. However, it is difficult to find any theoretical studies of cobalt ferrite films. The magnetic properties of cobalt ferrite films depend on film thickness [4].

Previously, the magnetic properties of thin nickel ferrite and ferromagnetic films have been explained by us using non-perturbed [6], $2^{nd}$ order perturbed [7, 8, 11] and $3^{rd}$ order perturbed [9, 10, 13] Heisenberg Hamiltonian modified by including $4^{th}$ order magnetic anisotropy and stress induced anisotropy. The variation of magnetic properties and easy axis orientation with number of layers and stress induced anisotropy of Nickel ferrite films have been described in those manuscripts. According to our previous studies, many magnetically easy and hard directions could be found with variations of stress induced anisotropy and number of layers. In this manuscript, the thick films of cobalt ferrite with thicknesses up to 10,000 unit cells have been considered, as the number of unit cells determines the thickness of the film. The spins of $Fe^{3+}$ and $Co^{2+}$ in the cell of cobalt ferrite have been taken into account for the simulations described in this manuscript. MATLAB computer software package was used for all the simulations. Stress induced anisotropy plays a major role in ferrite films. According to our previous experimental data, stress induced anisotropy plays a vital role in magnetic thin films [12, 14].

## 2. Model

Following modified Hamiltonian was used as the model.

$$H = -J\sum_{m,n}\vec{S}_m.\vec{S}_n + \omega\sum_{m\neq n}(\frac{\vec{S}_m.\vec{S}_n}{r_{mn}^3} - \frac{3(\vec{S}_m.\vec{r}_{mn})(\vec{r}_{mn}.\vec{S}_n)}{r_{mn}^5}) - \sum_m D_{\lambda_m}^{(2)}(S_m^z)^2 - \sum_m D_{\lambda_m}^{(4)}(S_m^z)^4$$

$$-\sum_m \vec{H}.\vec{S}_m - \sum_m K_s Sin2\theta_m \qquad (1)$$

Here J is spin exchange interaction, $\omega$ is the strength of long range dipole interaction, $\theta$ is azimuthal angle of spin, $D_m^{(2)}$ and $D_m^{(4)}$ are second and fourth order anisotropy constants, $H_{in}$ and $H_{out}$ are in plane and out of plane applied magnetic fields, $K_s$ is stress induced anisotropy constant, n and m are spin plane indices and N is total number of layers in film. When the stress applies normal to the film plane, the angle between $m^{th}$ spin and the stress is $\theta_m$.

The spinel cubic cell can be divided into 8 spin layers with alternative A and Fe spins layers [6]. The spins in one layer and adjacent layers point in one direction and opposite directions, respectively. The spins of A and Fe will be taken as 1 and p, respectively. A cubic unit cell with length a will be considered. Due to the super exchange interaction between spins, the spins are parallel or antiparallel to each other within the cell. Therefore the results proven for oriented case in one of our early report[6] will be used for following equations. But the angle $\theta$ will vary from $\theta_m$ to $\theta_{m+1}$ at the interface between two cells.

For a thin film with thickness Na,

Spin interaction energy=$E_{exchange}$= N(-10J+72Jp-22Jp$^2$)+8Jp$\sum_{m=1}^{N-1}\cos(\theta_{m+1} - \theta_m)$

Dipole interaction energy=$E_{dipole}$

$$E_{dipole} = -48.415\omega\sum_{m=1}^{N}(1+3\cos 2\theta_m) + 20.41\omega p\sum_{m=1}^{N-1}[\cos(\theta_{m+1}-\theta_m) + 3\cos(\theta_{m+1}+\theta_m)]$$

Here the first and second term in each above equation represent the variation of energy within the cell and the interface of the cell, respectively. Then total energy is given by

$$E(\theta) = N(-10J+72Jp-22Jp^2) + 8Jp\sum_{m=1}^{N-1}\cos(\theta_{m+1}-\theta_m)$$

$$-48.415\omega\sum_{m=1}^{N}(1+3\cos 2\theta_m) + 20.41\omega p\sum_{m=1}^{N-1}[\cos(\theta_{m+1}-\theta_m) + 3\cos(\theta_{m+1}+\theta_m)]$$

$$-\sum_{m=1}^{N}[D_m^{(2)}\cos^2\theta_m + D_m^{(4)}\cos^4\theta_m]$$

$$-4(1-p)\sum_{m=1}^{N}[H_{in}\sin\theta_m + H_{out}\cos\theta_m + K_s\sin 2\theta_m] \qquad (2)$$

Here the anisotropy energy term and the last term have been explained in our previous report for oriented spinel ferrite [6]. If the angle is given by $\theta_m=\theta+\varepsilon_m$ with perturbation $\varepsilon_m$, after taking the terms up to third order perturbation of $\varepsilon$,

The total energy can be given as $E(\theta)=E_0+E(\varepsilon)+E(\varepsilon^2)+E(\varepsilon^3)$

Here

$E_0 = -10JN+72pNJ-22Jp^2N+8Jp(N-1)-48.415\omega N-145.245\omega N\cos(2\theta)$
$\qquad +20.41\omega p[(N-1)+3(N-1)\cos(2\theta)]$

$$-\cos^2\theta\sum_{m=1}^{N}D_m^{(2)} - \cos^4\theta\sum_{m=1}^{N}D_m^{(4)} - 4(1-p)N(H_{in}\sin\theta + H_{out}\cos\theta + K_s\sin 2\theta) \qquad (3)$$

$$E(\varepsilon) = 290.5\omega\sin(2\theta)\sum_{m=1}^{N}\varepsilon_m - 61.23\omega p\sin(2\theta)\sum_{m=1}^{N-1}(\varepsilon_m+\varepsilon_n)$$

$$+\sin 2\theta\sum_{m=1}^{N}D_m^{(2)}\varepsilon_m + 2\cos^2\theta\sin 2\theta\sum_{m=1}^{N}D_m^{(4)}\varepsilon_m$$

$$+4(1-p)[-H_{in}\cos\theta\sum_{m=1}^{N}\varepsilon_m + H_{out}\sin\theta\sum_{m=1}^{N}\varepsilon_m - 2K_s\cos 2\theta\sum_{m=1}^{N}\varepsilon_m] \qquad (4)$$

$$E(\varepsilon^2) = -4Jp\sum_{m=1}^{N-1}(\varepsilon_n - \varepsilon_m)^2 + 290.5\omega\cos(2\theta)\sum_{m=1}^{N}\varepsilon_m^2 - 10.2\omega p\sum_{m=1}^{N-1}(\varepsilon_n - \varepsilon_m)^2$$

$$- 30.6\omega p\cos(2\theta)\sum_{m=1}^{N-1}(\varepsilon_n + \varepsilon_m)^2$$

$$+ \cos 2\theta\sum_{m=1}^{N}D_m^{(2)}\varepsilon_m^2 + 2\cos^2\theta(\cos^2\theta - 3\sin^2\theta)\sum_{m=1}^{N}D_m^{(4)}\varepsilon_m^2$$

$$+ 4(1-p)[\frac{H_{in}}{2}\sin\theta\sum_{m=1}^{N}\varepsilon_m^2 + \frac{H_{out}}{2}\cos\theta\sum_{m=1}^{N}\varepsilon_m^2 + 2K_s\sin 2\theta\sum_{m=1}^{N}\varepsilon_m^2] \quad (5)$$

$$E(\varepsilon^3) = 10.2 p\omega\sin 2\theta\sum_{m,n=1}^{N}(\varepsilon_m + \varepsilon_n)^3 - 193.66\omega\sin 2\theta\sum_{m=1}^{N}\varepsilon_m^3 - \frac{4}{3}\cos\theta\sin\theta\sum_{m=1}^{N}D_m^{(2)}\varepsilon_m^3$$

$$- 4\cos\theta\sin\theta(\frac{5}{3}\cos^2\theta - \sin^2\theta)\sum_{m=1}^{N}D_m^{(4)}\varepsilon_m^3$$

$$+ 4(1-p)[\frac{H_{in}}{6}\cos\theta\sum_{m=1}^{N}\varepsilon_m^3 - \frac{H_{out}}{6}\sin\theta\sum_{m=1}^{N}\varepsilon_m^3 + \frac{4K_s}{3}\cos 2\theta\sum_{m=1}^{N}\varepsilon_m^3]$$

The sin and cosine terms in equation number 2 have been expanded to obtain above equations. Here n=m+1.

Under the constraint $\sum_{m=1}^{N}\varepsilon_m = 0$, first and last three terms of equation 4 are zero.

Therefore, $E(\varepsilon) = \vec{\alpha}.\vec{\varepsilon}$

Here $\vec{\alpha}(\varepsilon) = \vec{B}(\theta)\sin 2\theta$ are the terms of matrices with

$$B_\lambda(\theta) = -122.46\omega p + D_\lambda^{(2)} + 2D_\lambda^{(4)}\cos^2\theta \quad (6)$$

Also $E(\varepsilon^2) = \frac{1}{2}\vec{\varepsilon}.C.\vec{\varepsilon}$, and matrix C is assumed to be symmetric ($C_{mn}=C_{nm}$).

Here the elements of matrix C can be given as following,

$C_{m, m+1}$=8Jp+20.4ωp-61.2pωcos(2θ)

For m=1 and N,

$C_{mm}$= -8Jp-20.4ωp-61.2pωcos(2θ)+581ωcos(2θ) + 2cos 2θ $D_m^{(2)}$

$$+ 4\cos^2\theta(\cos^2\theta - 3\sin^2\theta) D_m^{(4)} + 4(1-p)[H_{in}\sin\theta + H_{out}\cos\theta + 4K_s\sin(2\theta)] \quad (7)$$

For m=2, 3, ----, N-1

$$C_{mm}= -16J_p - 40.8\omega p - 122.4 p\omega \cos(2\theta) + 581\omega \cos(2\theta) + 2\cos 2\theta\, D_m^{(2)}$$

$$+ 4\cos^2\theta(\cos^2\theta - 3\sin^2\theta)\, D_m^{(4)} + 4(1-p)[H_{in}\sin\theta + H_{out}\cos\theta + 4K_s\sin(2\theta)]$$

Otherwise, $C_{mn}=0$

Also $E(\varepsilon^3) = \varepsilon^2 \beta.\vec{\varepsilon}$

Here matrix elements of matrix $\beta$ can be given as following.

When m=1 and N,

$$\beta_{mm} = -193.66\omega\sin 2\theta + 10.2 p\omega\sin 2\theta - \frac{4}{3}\cos\theta\sin\theta D_m^{(2)}$$

$$- 4\cos\theta\sin\theta(\frac{5}{3}\cos^2\theta - \sin^2\theta)D_m^{(4)} + 4(1-p)[\frac{H_{in}}{6}\cos\theta - \frac{H_{out}}{6}\sin\theta + \frac{4K_s}{3}\cos 2\theta]$$

When m=2, 3, ------, N-1

$$\beta_{mm} = -193.66\omega\sin 2\theta + 20.4 p\omega\sin 2\theta - \frac{4}{3}\cos\theta\sin\theta D_m^{(2)}$$

$$- 4\cos\theta\sin\theta(\frac{5}{3}\cos^2\theta - \sin^2\theta)D_m^{(4)} + 4(1-p)[\frac{H_{in}}{6}\cos\theta - \frac{H_{out}}{6}\sin\theta + \frac{4K_s}{3}\cos 2\theta]$$

$$\beta_{m,m+1} = 30.6 p\omega\sin 2\theta \tag{8}$$

Otherwise $\beta_{nm}=0$. Also matrix $\beta$ is symmetric such that $\beta_{nm}=\beta_{mn}$.

Finally, the total magnetic energy given in equation 2 can be deduced to

$$E(\theta)=E_0 + \vec{\alpha}.\vec{\varepsilon} + \frac{1}{2}\vec{\varepsilon}.C.\vec{\varepsilon} + \varepsilon^2 \beta.\vec{\varepsilon} \tag{9}$$

Only the second order terms of $\varepsilon$ will be considered for following derivation, since the derivation with the third order terms of $\varepsilon$ in above equation is tedious.

Then $E(\theta)=E_0 + \vec{\alpha}.\vec{\varepsilon} + \frac{1}{2}\vec{\varepsilon}.C.\vec{\varepsilon}$

Using a suitable constraint in above equation, it is possible to show that $\vec{\varepsilon} = -C^+.\vec{\alpha}$

Here $C^+$ is the pseudo-inverse given by

$$C.C^+ = 1 - \frac{E}{N}. \qquad (10)$$

Each element in matrix $E$ is 1.

After using $\vec{\varepsilon}$ in equation 9, $E(\theta) = E_0 - \frac{1}{2}\vec{\alpha}.C^+.\vec{\alpha} - (C^+\alpha)^2 \vec{\beta}(C^+\alpha)$ (11)

## 3. Results and Discussion

$C^+$ given in equation 10 will be deduced to the standard inverse matrix of C, when N is really large. When the difference between m and n is one, $C_{m, m+1} = 8Jp + 20.4\omega p - 61.2 p\omega \cos(2\theta)$. If $H_{in}$, $H_{out}$ and $K_s$ are very large, then $C_{11} \gg C_{12}$. If this $C_{m, m+1} = 0$, then the matrix C becomes diagonal, and the elements of inverse matrix $C^+$ is given by $C^+_{mm} = \frac{1}{C_{mm}}$. Therefore all the derivation will be done under above assumption to avoid tedious derivations. Value of p for cobalt ferrite is 1.67.

From equation 3,

$E_0 = -10JN + 120.24NJ - 61.36JN + 13.36J(N-1) - 48.415\omega N - 145.245\omega N\cos(2\theta)$
$\quad + 34.1\omega[(N-1) + 3(N-1)\cos(2\theta)]$

From equation 7,

$C_{11} = C_{NN} = -13.36J - 34.07\omega + 478.8\omega\cos(2\theta) + 2\cos 2\theta\, D_m^{(2)}$
$\quad + 4\cos^2\theta(\cos^2\theta - 3\sin^2\theta)\, D_m^{(4)} - 2.68[H_{in}\sin\theta + H_{out}\cos\theta + 4K_s\sin(2\theta)]$

For m=2, 3, ----, N-1

$C_{22} = C_{33} = ---- = C_{N-1,N-1} = -26.72J - 68.14\omega + 376.59\omega\cos(2\theta) + 2\cos 2\theta\, D_m^{(2)}$
$\quad + 4\cos^2\theta(\cos^2\theta - 3\sin^2\theta)\, D_m^{(4)} - 2.68[H_{in}\sin\theta + H_{out}\cos\theta + 4K_s\sin(2\theta)]$

From equation 6,

$\alpha_1 = [-306.15\omega + D_\lambda^{(2)} + 2D_\lambda^{(4)}\cos^2\theta]\sin(2\theta)$

$\beta_{11} = \beta_{NN} = -168.16\omega\sin 2\theta - \frac{4}{3}\cos\theta\sin\theta D_m^{(2)}$

$\quad - 4\cos\theta\sin\theta(\frac{5}{3}\cos^2\theta - \sin^2\theta)D_m^{(4)} - 6[\frac{H_{in}}{6}\cos\theta - \frac{H_{out}}{6}\sin\theta + \frac{4K_s}{3}\cos 2\theta]$

$$\beta_{22} = -142.66\omega\sin 2\theta - \frac{4}{3}\cos\theta\sin\theta D_m^{(2)}$$

$$-4\cos\theta\sin\theta(\frac{5}{3}\cos^2\theta - \sin^2\theta)D_m^{(4)} - 6[\frac{H_{in}}{6}\cos\theta - \frac{H_{out}}{6}\sin\theta + \frac{4K_s}{3}\cos 2\theta]$$

$$\beta_{m,m+1} = 76.5\omega\sin 2\theta$$

$(C^+\alpha)^2\beta(C^+\alpha) = (C_{11}^+\alpha_1)^2(\beta_{11}C_{11}^+\alpha_1 + \beta_{12}C_{22}^+\alpha_2 + \text{-------} + \beta_{1N}C_{NN}^+\alpha_N)$

$\qquad + (C_{22}^+\alpha_2)^2(\beta_{21}C_{11}^+\alpha_1 + \beta_{22}C_{22}^+\alpha_2 + \text{-------} + \beta_{2N}C_{NN}^+\alpha_N)$

$\qquad + (C_{33}^+\alpha_3)^2(\beta_{31}C_{11}^+\alpha_1 + \beta_{32}C_{22}^+\alpha_2 + \text{-------} + \beta_{3N}C_{NN}^+\alpha_N) + \text{--------}$

$\qquad \text{------} + (C_{NN}^+\alpha_N)^2(\beta_{N1}C_{11}^+\alpha_1 + \beta_{N2}C_{22}^+\alpha_2 + \text{-------} + \beta_{NN}C_{NN}^+\alpha_N)$

$$(C^+\alpha)^2\beta(C^+\alpha) = \alpha^3[\frac{2}{C_{11}^2}(\frac{\beta_{11}}{C_{11}} + \frac{\beta_{12}}{C_{22}}) + \frac{2}{C_{22}^2}(\frac{\beta_{12}}{C_{11}} + \frac{\beta_{22} + \beta_{12}}{C_{22}}) + \frac{N-4}{C_{22}^3}(2\beta_{12} + \beta_{22})]$$

The total magnetic energy can be found by putting all these terms in equation 11.

Figure 1 is the 3-D plot of energy versus in plane magnetic field and angle for $\frac{J}{\omega} = \frac{D_m^{(2)}}{\omega} = \frac{H_{out}}{\omega} = \frac{K_s}{\omega} = 10$ and $\frac{D_m^{(4)}}{\omega} = 5$. The number of layers was taken as 10,000 in this case. Energy minimums can be observed at $\frac{H_{in}}{\omega} = 10, 13, 28, 56, 69, \text{---etc}$, indicting that the film can be easily oriented along easy direction of magnetization at these values of in plane magnetic field. Also the graph has maximum values at $\frac{H_{in}}{\omega} = 5, 30, 36, 50, 55, 61, \text{----etc}$. This means that it is difficult to align the film in hard directions at these values of $\frac{H_{in}}{\omega}$. However, there are several easy and hard directions at one of these values of in plane magnetic fields.

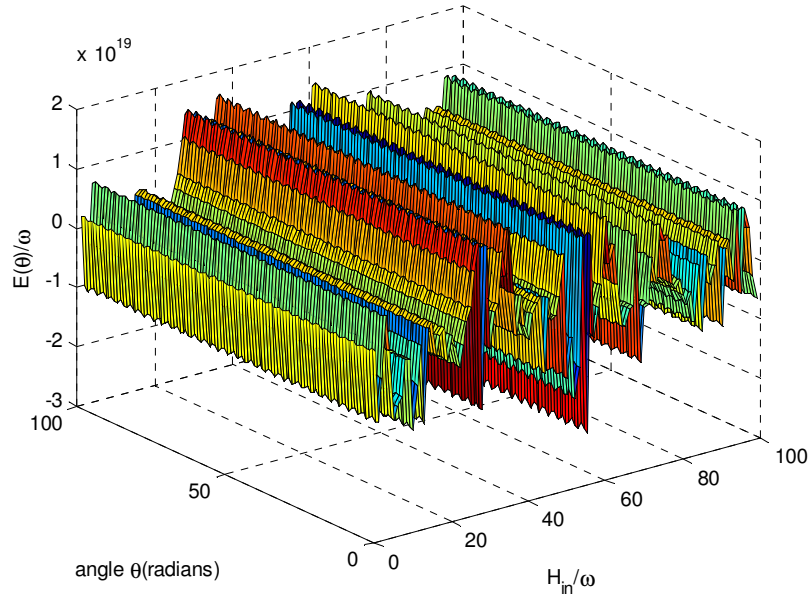

*Fig 1: Graph of energy versus angle and in plane magnetic field.*

2-D plot in figure 2 shows the variation of total magnetic energy with angle at $\frac{H_{in}}{\omega} = 28$. Here, the other parameters were kept at $\frac{J}{\omega} = \frac{D_m^{(2)}}{\omega} = \frac{H_{out}}{\omega} = \frac{K_s}{\omega} = 10$ and $\frac{D_m^{(4)}}{\omega} = 5$. Number of layers was kept at 10,000. Easy axis can be observed at 3.4243 and 5.8434 radians for this particular value of in plane magnetic field.

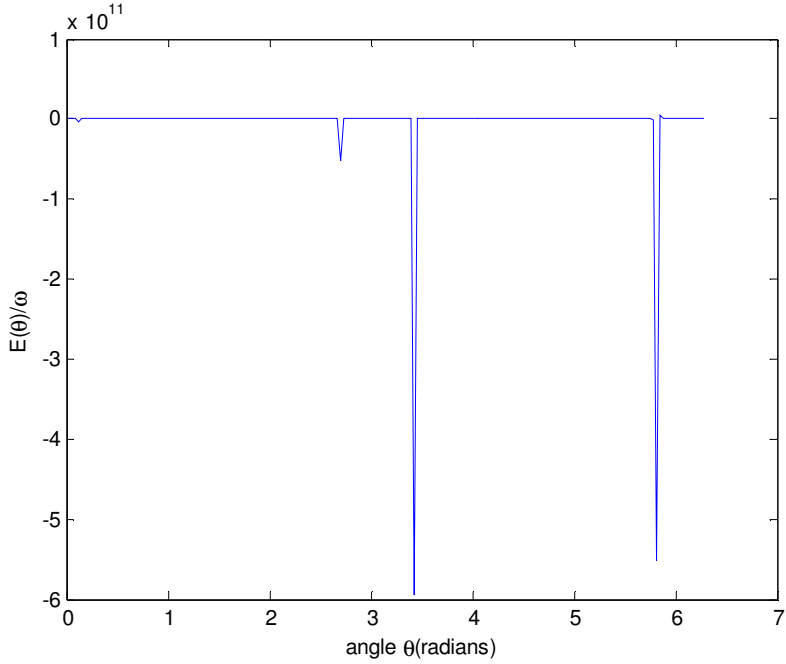

**Fig 2:** *Graph of energy versus angle at* $\frac{H_{in}}{\omega} = 28$.

The total magnetic energy versus out of plane magnetic field and angle was plotted in figure 3. In this case, the other parameters were kept at $\frac{J}{\omega} = \frac{D_m^{(2)}}{\omega} = \frac{H_{in}}{\omega} = \frac{K_s}{\omega} = 10$ and $\frac{D_m^{(4)}}{\omega} = 5$. Number of layers was kept at 10,000. There are energy minimums in the graph at $\frac{H_{out}}{\omega} = 15, 23, 37, 49, 72,$ --- etc. This means that the film can be easily oriented along magnetically easy direction at these values of out of plane magnetic fields. Energy maximums can be observed at $\frac{H_{out}}{\omega} = 10, 16, 20, 42, 64,$ ---- etc. It is difficult to align the film in hard directions at these values of $\frac{H_{out}}{\omega}$.

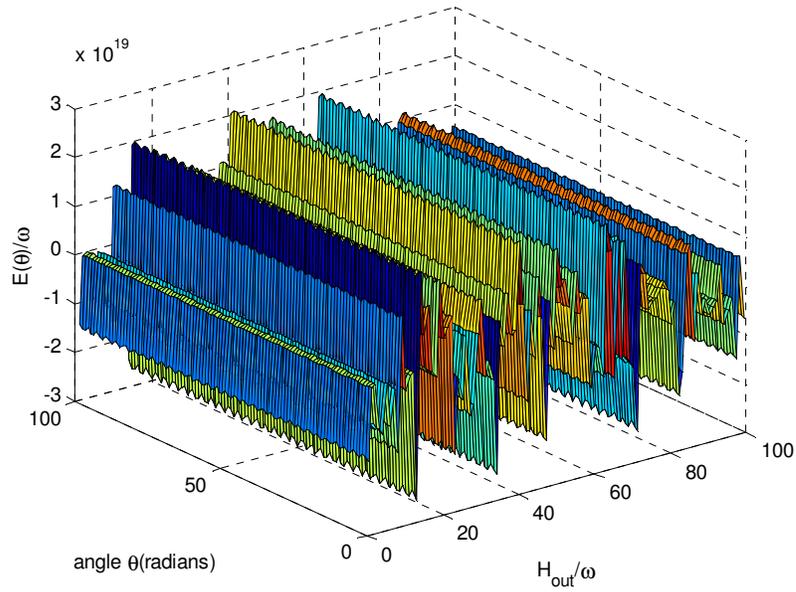

*Fig 3: Graph of energy versus angle and out of plane magnetic field.*

In figure 4, the graph of energy versus angle has been plotted for $\frac{H_{out}}{\omega} = 37$. Other parameters were kept at the values given for figure 3. Easy directions make 0.4084 and 6.15 radians with a normal line drawn to film plane at this specific value of out of magnetic plane.

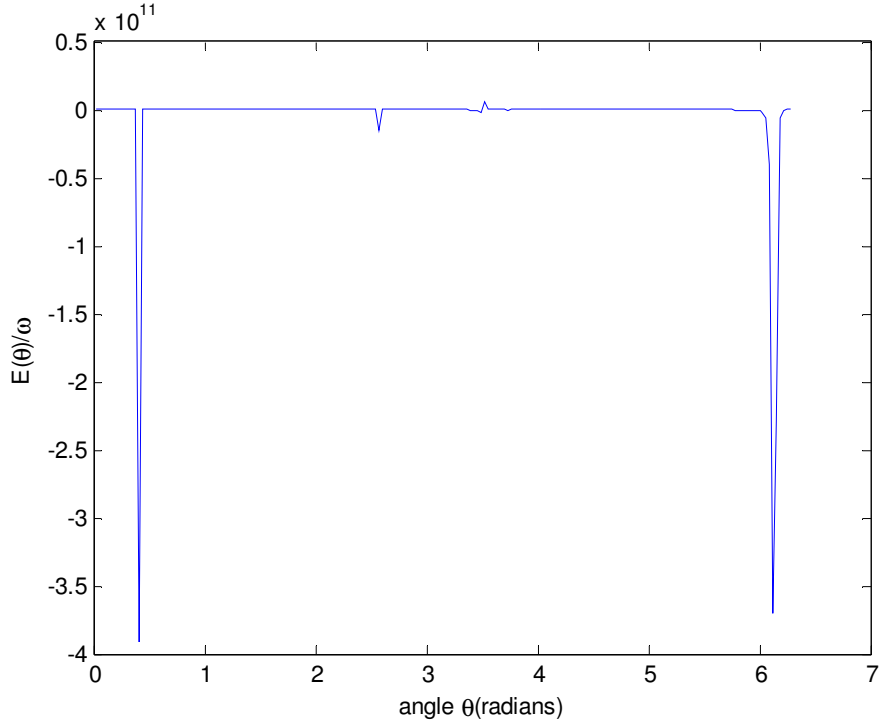

**Fig 4:** *Graph of energy versus angle at* $\dfrac{H_{out}}{\omega} = 37$.

## 4. Conclusion

Easy and hard directions of cobalt ferrite films at different values of in plane and out of plane magnetic fields have been investigated. Easy direction as measured with normal line drawn to film plane is 196.2 and 334.8 degrees at N=10,000, $\dfrac{H_{in}}{\omega} = 28$, $\dfrac{J}{\omega} = \dfrac{D_m^{(2)}}{\omega} = \dfrac{H_{out}}{\omega} = \dfrac{K_s}{\omega} = 10$ and $\dfrac{D_m^{(4)}}{\omega} = 5$. Also easy directions were found to be 23.4 and 352.4 degrees for N=10,000, $\dfrac{H_{out}}{\omega} = 37$, $\dfrac{J}{\omega} = \dfrac{D_m^{(2)}}{\omega} = \dfrac{H_{in}}{\omega} = \dfrac{K_s}{\omega} = 10$ and $\dfrac{D_m^{(4)}}{\omega} = 5$. As given in 3-D plots, easy and hard directions are determined by spin exchange interaction, dipole interaction, second & fourth order anisotropy, in plane & out of plane magnetic fields, stress induced anisotropy and the thickness of the film. Therefore easy and hard directions of thick cobalt ferrite films can be tuned by varying these energy parameters.

# REFERENCES

1. L. Stichauer *et al.,* 1996. Optical and magneto-optical properties of nanocrystalline cobalt ferrite films. Journal of Applied Physics 79, 3645.

2. ZhiYong Zhong *et al.,* 2011. Microstructure and magnetic properties of $CoFe_2O_4$ thin films deposited on Si substrates with an $Fe_3O_4$ under layer. Science China: Physics, Mechanics and Astronomy 54 (7), 1235-1238.

3. N. Hiratsuka and M. Sugimoto, 1987. Preparation of amorphous cobalt ferrite films with perpendicular anisotropy and their magneto optical properties. IEEE Transaction on Magnetism MAG. 23(5), 3326-3328.

4. Subasa C. Sahoo *et al.,* 2012. Thickness dependent anomalous magnetic behavior in pulsed laser deposited cobalt ferrite thin films. Applied Physics A 106(4), 931-935.

5. G. Dascalu *et al,* 2013. Magnetic measurements of RE doped cobalt ferrite thin films. IEEE Transaction on Magnetism 49 (1), 46-49.

6. P. Samarasekara, 2007. Classical Heisenberg Hamiltonian Solution of Oriented Spinel Ferrimagnetic Thin Films. Electronic Journal of Theoretical Physics 4(15), 187-200.

7. P. Samarasekara, M.K. Abeyratne and S. Dehipawalage, 2009. Heisenberg Hamiltonian with Second Order Perturbation for Spinel Ferrite Thin Films. Electronic Journal of Theoretical Physics 6(20), 345-356.

8. P. Samarasekara, 2010. Determination of energy of thick spinel ferrite films using Heisenberg Hamiltonian with second order perturbation. Georgian electronic scientific journals: Physics 1(3), 46-49.

9. P. Samarasekara and William A. Mendoza, 2011. Third Order Perturbed Heisenberg Hamiltonian of Spinel Ferrite Ultra-thin films. Georgian Electronic Scientific Journals: Physics 1(5), 15-24.

10. P. Samarasekara, 2011. Investigation of Third Order Perturbed Heisenberg Hamiltonian of Thick Spinel Ferrite Films. Inventi Rapid: Algorithm Journal 2(1), 1-3.

11. P. Samarasekara and S.N.P. De Silva, 2007. Heisenberg Hamiltonian solution of thick ferromagnetic films with second order perturbation. Chinese Journal of Physics 45(2-I), 142-150.

12. P. Samarasekara, 2003. A pulsed rf sputtering method for obtaining higher deposition rates. Chinese Journal of Physics 41(1), 70-74.
13. P. Samarasekara and William A. Mendoza, 2010. Effect of third order perturbation on Heisenberg Hamiltonian for non-oriented ultra-thin ferromagnetic films. Electronic Journal of Theoretical Physics 7(24), 197-210.
14. P. Samarasekara and Udara Saparamadu, 2013. Easy axis orientation of Barium hexa-ferrite films as explained by spin reorientation. Georgian electronic scientific journals: Physics 1(9), 10-15.